\documentclass[12pt]{iopart}
\usepackage{color}
\definecolor{darkblue}{rgb}{0.,0.,0.4}

\usepackage{iopams,graphicx,bm}
\usepackage[breaklinks=true,colorlinks=true,linkcolor=blue,urlcolor=darkblue,citecolor=blue]{hyperref}
\usepackage[numbers,sort&compress]{natbib}
\begin{document}

\title[A Nanographene Disk Rotating a single Molecule Gear on a Cu(111) Surface]{A Nanographene Disk Rotating a single Molecule Gear on a Cu(111) Surface}

\author{H.-H. Lin$^{1}$, A. Croy$^1$, R. Gutierrez$^1$, C. Joachim$^2$ and G. Cuniberti$^{1}$}

\address{$^1$
 Institute for Materials Science and Max Bergmann Center of Biomaterials, TU Dresden, 01069 Dresden, Germany\\
}
\address{$^2$
 GNS and MANA Satellite, CEMES-CNRS, 29 rue J. Marvig, 31055 Toulouse Cedex, France\\
}
\eads{\mailto{alexander.croy@tu-dresden.de}, \mailto{gianaurelio.cuniberti@tu-dresden.de}}
\vspace{10pt}
\begin{indented}
\item[]August 2021
\end{indented}

\begin{abstract}
We perform molecular dynamics simulations to study the collective rotation of a graphene nanodisk functionalized on its circumference by \textit{tert}-butylphenyl chemical groups in interaction with a molecule-gear hexa-\textit{tert}-butylphenylbenzene supported by a Cu(111) surface. The rotational motion can be categorized under-driving, driving and overdriving regimes calculating the locking coefficient of this machinery as a function of external torque applied. Moreover, the rotational friction with the surface of both the phononic and electronic contributions is investigated. It shows that for small size graphene nanodisks the phononic friction is the main contribution, whereas the electronic one dominates for the larger disks putting constrains on the experimental way of achieving the transfer of rotation from a graphene nanodisk to single molecule-gear.
\end{abstract}

%
%
%
%
%

\section{\label{sec:Introduction}Introduction}

Progress in the miniaturization of solid states gears down to the nanoscale\cite{JuYun2007, Yang2014,Mailly2020} is calling for the study of gear trains and of the transfer of rotation from a given gear diameter to the next (up and down in scale) \cite{Deng2011}. We are now reaching the experimental stage where a single molecule-gear 1 nm in diameter\cite{Manzano2009} will have to be rotated by the smallest possible solid state nanogear with diameter in the 10 nm range\cite{Yang2014,Mailly2020}. Already the transfer of rotation from  one molecule-gear to the next has been observed along a molecule-gear train where each rotation axle along the train is a single metallic ad-atom\cite{WeiHyo2019}. The nanotechnology handicraft required to construct a nanoscale machinery to transfer rotation from a solid state nanogear to a molecule is under active exploration\cite{Joachim2020a}. It encompasses gear thickness compatibility, atomic scale surface preparation and preservation after transferring the solid state nanogear on it, rotation axle stability for both the solid state and the molecule gears and the means to activate the rotation of the solid state nanogear to observe the rotation of the molecule-gear. There is here also the fundamental question of the interactions of both the solid state nanogear and the molecule-gear of this machinery with the supporting surface and how this impacts the torque required to rotate both gears starting from the solid state one.

\begin{figure}[t]
\centering
 \includegraphics[width=0.8\textwidth]{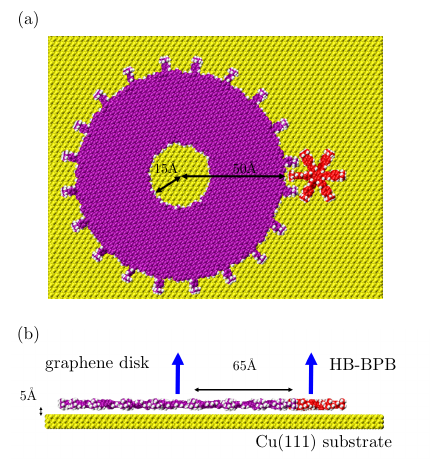}
 \caption{A schematic illustration of the (a) topview (b) sideview for a \textit{tert}-butylbiphenyl functionalized graphene disk with radius and hexa-\textit{tert}-butylbiphenylbenzene on Cu(111) surface.}
 \label{Fig:setup}
\end{figure}
 In this work, we present a theoretical study of the transmission of rotation between a single layer  graphene nanogear and a single molecule-gear both adsorbed on the same atomically defined surface (Fig.~\ref{Fig:setup}). The molecule-gear is an hexa-\textit{tert} -butylphenylbenzene (HB- BPB) molecule with  6 \textit{tert}-butylphenyl teeth\cite{Soe2020}. The  10 nm in diameter graphene nanodisk is supposed to be also equipped with \textit{tert}-butylphenyl teeth. With this design, the chemical structure of the teeth on both gears is identical. The thickness of the graphene nanodisk (hereafter called the master) is exactly compatible with the molecule-gear chemical structure and with its physisorption height on the Cu(111) surface. This compatibility is usually difficult to reach experimentally since electron beam nanolithography processes ultimately lead to a 5 to 10 nm gear thickness\cite{Yang2014,Mailly2020}. A newcomer in the clean room, the focused He$^+$ beam nanolithography (HIM) will normally lead to the possibility of sculpturing one after the other graphene monolayer nanodisks in high vacuum\cite{WeiHyo2019}.

With a circumference nanodisk edge having essentially sp$^2$ carbon like dangling bonds at the end of the HIM process, this is opening the way to chemically bond the \textit{tert}-butylphenyl teeth also on the graphene nanodisk. As presented in Fig.~\ref{Fig:setup} and for a 10 nm diameter graphene nanodisk master, we have determined the optimal number of \textit{tert}-butylphenyl around its circumference to ensure a nice transmission of rotation. We found that 20 \textit{tert}-butylpheny chemical groups are working well.  Other master diameters will certainly require to adapt again its edge chemical structure. There is also a central 1.5 nm hole in the master Fig.~\ref{Fig:setup} to figure out the location of the rotation axle which for example can be a mold of copper ad-atoms. The center-of-mass separation between this master and the HB-BPB molecule-gear is 6.5 nm. This was also optimized for optimum transmission of rotation. We have already observed experimentally the importance of the pm precision of this center of mass separation relative to the chemical structure of the teeth\cite{Soe2020,WeiHyo2019}. To model the Cu(111) supporting surface, we consider hereafter a three layers Cu(111) surface (with size 15.79 nm $\times$ 12.766 nm $\times$ 0.23 nm) made of 10080 atoms with periodic boundary conditions in the lateral $x$ and $y$ directions. We first preposition the master and its HB-BPB molecule-gear 0.5 nm above the Cu(111) surface.  To avoid a net translation of this surface after the application of a torque to the master (action-reaction principle), we fix the bottom Cu layer and allow only to relax the upper two layers. Finally, and before starting the calculations, a geometry optimization of the complete Fig.~\ref{Fig:setup} structure was performed using the conjugate gradient method to start with an optimized surface conformation for the Fig.~\ref{Fig:setup} machinery.

We first demonstrated how with the Fig.~\ref{Fig:setup} design, a 5 nm graphene nanodisk can effectively transfer a regular rotation movement to the interacting HB-BPB molecule-gear. This is functioning in a restricted range of torque applied to the master required to fight against the surface friction of the master but not to destabilize the molecule-gear chemical structure itself by using a too large torque. In a second step, we present a more detail theoretical study of the effect of mechanical surface friction as a function of the master diameter and compare it with its electronic friction on the Cu(111) surface. In conclusion, we discuss how the  rather simple nanoscale mechanical machinery of Fig.~\ref{Fig:setup} can be operated experimentally.

\section{\label{sec:Formalism} Model system and computational approach }
In this section, we introduce the details concerning our molecular dynamic (MD) simulations, describe the near rigid-body approximation used and recall the concept  of  locking  coefficient well known at the macroscopic scale to study gearing effects along a train of gears.  To carry out the MD simulations on the  nanoscale machinery of Fig.~\ref{Fig:setup}, we exploit the Large-scale Atomic/Molecular Massively Parallel Simulator (LAMMPS)\cite{Plimpton1995}. For the force fields, we choose the Reactive force field (ReaxFF)\cite{Plimpton1995} parametrized for Cu, C and H. To fix the rotational axles, we connect a stiff spring with spring constant $k=1600$ N/m (1000 eV/\AA$^2$) to each gear center-of-mass. In this way  we avoid the atomic construction of the axles especially of the Cu ad-atoms cluster mold, for the master axle of rotation is not essential at this stage. To determine the temperature of the supporting surface, the upper two Cu layers are subject to the canonical ensemble implemented here by a Nos\'e–Hoover thermostat\cite{Nose1984,Hoover1985} reaching a surface temperature $T$  = 10K (the actual single molecule mechanics experiments with a low temperature scanning tunneling microscope (STM) are performed at about 5 K\cite{Manzano2009,WeiHyo2019}). To extract from the MD simulation the different conformation angles of the deformable master nanodisk and of the molecule-gear, the nearly rigid-body approximation\cite{Lin2019a, Lin2020} was exploited. For both the master and the molecule-gear in Fig.~\ref{Fig:setup} , we define the reference atomic scale structures by a set of coordinates $\{{\bm{r}^0_{ k_{\alpha}}}\}$, where $k_{\alpha}$ runs through all atoms of the gear $k_{\alpha}$ with $\alpha=1,2$ for the master and the molecule-gear, respectively. For example, we can choose the reference atomic structure to be the one at time $t=t_0$ with its optimized coordinates for the master and molecule-gear before performing the MD simulation $\{{\bm{r}_{k_{\alpha}}(t_0)}\}$.   Supposing that the coordinates at time $t$ is given by \{${\bm{r}_{\alpha k}(t)}$\}, the deformation in all frames are here assumed to be sufficiently small such that the structure \{${\bm{r}_{\alpha k}(t)}$\} can still be mapped to the reference atomic structure $\{{\bm{r}_{k_{\alpha}}(t_0)}\}$ via a rigid-body rotational transformation matrix $\bm{R}^{\alpha}(\theta_{\alpha }(t),\bm{n}_{\alpha}(t))$ defining the rotational axes $\bm{n}_{\alpha}(t)$ and the angles $\theta_{\alpha}(t)$. The rigid-body transformation matrix $\bm{R}^{\alpha}(\theta_{\alpha }(t),\bm{n}_{\alpha}(t))$ can be found by minimizing the deformation using the quaternions method\cite{Kneller1991,Lin2020}. To characterize the mechanical transmission of rotation between the master and the molecule-gear, we use the concept of locking coefficient\cite{Lin2020}, well known for macroscopic gears, and defined here by:
\begin{equation}
    L_{\alpha}=\frac{\langle \omega_{\alpha} \rangle}{\omega_{R\alpha}}\;,
    \label{eq:Locking}
\end{equation}
where $\langle \omega_{\alpha}\rangle$ denotes the average angular velocity of gear $\alpha$ with $\omega_{R\alpha}$ the terminal angular velocity of the rigid-body corresponding to this gear $\alpha$. This quantity provides a measure of the ability to transfer a rotation between the two gears in Fig.~\ref{Fig:setup}. Note that, for perfectly interlocked rigid gears, the coefficients $L_{\alpha}$ must be equal to unity. For soft gears or a not perfect transmission, we expect $L_{\alpha}$ to be much smaller than unity as discussed below.

\begin{figure}[t]\centering
 \includegraphics[width=0.6\textwidth]{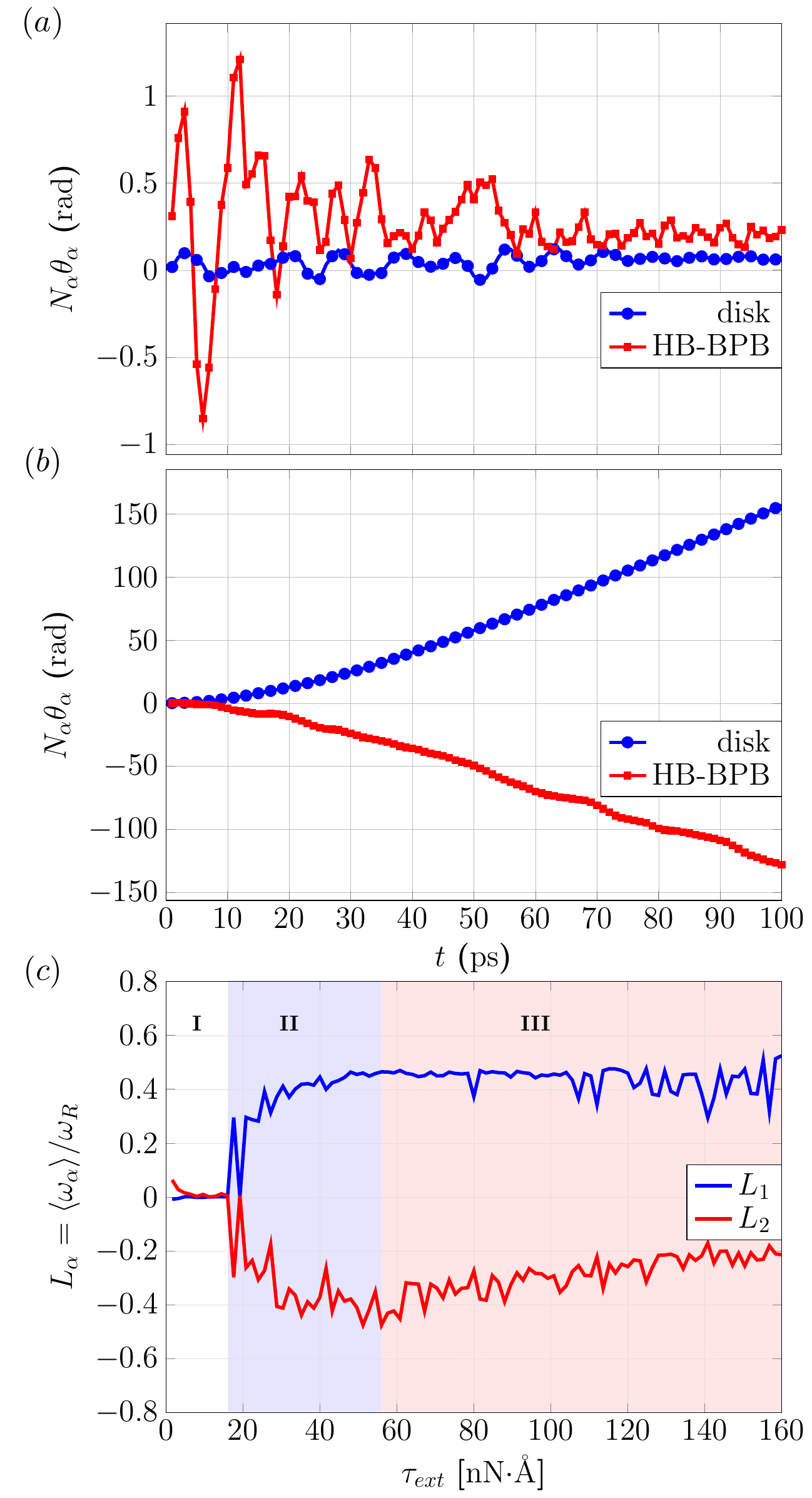}
 \caption{The angle displacement multiplied with number of teeth $N_1=20$ for graphene disk ($N_2=6$ for HB-BPB) within 100 ps with external torque (a) $\tau_{ext}=12.8$ nN$\cdot$\AA{} and (b) $\tau_{ext}=48$  nN$\cdot$\AA{}. The blue (red) line represents the. (c) The locking coefficient $L_{1,2}$ with respect to external torque ranged from 0 to 160  nN$\cdot$\AA{}, where region I, II and III represent underdriving, driving and overdriving phases, respectively.}
 \label{Fig:disk_pinion_Cu111_trajectory}
\end{figure}
\section{\label{sec:results} Results and discussions}
\subsection{Rotational transmission}
To investigate the transmission of rotation in the machinery shown in Fig.~\ref{Fig:setup}, we apply an external torque $\tau_{ext}$ on the master and follow how the corresponding angular moment is transferred (or not) from this master to the HB-BPB molecule. The $N_{\alpha}\theta_{\alpha}(t)$ variations in time for a small $\tau_{ext}=12.8$ nN$\cdot$\AA{} and an intermediate  $\tau_{ext}=48$ nN$\cdot$\AA{}  torques  are  presented  in Fig.~\ref{Fig:disk_pinion_Cu111_trajectory} (a) and (b) with $N_1= 20$ and $N_2 = 6$ respectively. For a small torque (Fig.~\ref{Fig:disk_pinion_Cu111_trajectory} (a)), we found only small amplitude oscillations of both gears around their equilibrium conformation. The applied torque is not large enough to fight against the master surface friction on Cu(111) and to permit the molecule-gear to pass over the energy barrier of the potential energy surface on Cu(111) at a $T$= 10 K surface temperature. Notice also the damping effect of the Cu(111) surface is more effective to reduce the oscillation amplitudes of the master as compared to the ones of the molecule-gear (see also section III.B below). On the other hand, for the selected intermediate torque in Fig.~\ref{Fig:disk_pinion_Cu111_trajectory}(b), we found a perfectly interlocked rotation respecting the classical condition $N_1\theta_1+N_2\theta_2=0$\cite{MacKinnon2002}. Let us recall here that to reach such a result, we have equipped the edge of the master with exactly 20 equi-spaced \textit{tert}-butylpheny chemical groups. To understand this perfect transmission of rotation, we have also plotted the locking coefficient as a function of torque in Fig.~\ref{Fig:disk_pinion_Cu111_trajectory}(c). To calculate Eq. \ref{eq:Locking} for the master, we have calculated its terminal velocity $\omega_{R1}$ using:
\begin{equation}
	\omega_{R1}=\frac{\tau_{ext}}{\gamma_1},
\end{equation}
where $\gamma_1=2.83\times 10^{-29}$ kg$\cdot$m$^2\cdot$s$^{-1}$ (extracted from $\tau_{ext}/\langle\omega_{1}\rangle$ in the same MD simulation) is the damping coefficient corresponding to rigid disk on the Cu surface. For $\omega_{R2}$, we can take the time derivative of the interlocked condition $N_1\theta_1+N_2\theta_2=0$, leading to
\begin{equation}
	\omega_{R2}=-\frac{N_1\omega_1}{N_2}.
\end{equation}
The resulting locking coefficients are presented in Fig.~\ref{Fig:disk_pinion_Cu111_trajectory} (c) where three different regimes can be distinguished for the transmission of rotation from the master to the molecule-gear.  In region I (white) for $\tau_{ext}<18$ nN$\cdot$\AA{}, $|L_1|\approx |L_2|$. In this case, the average angular velocity for both the master and the molecule-gear are vanishing. There is no rotation. It is the so-called underdriving regime\cite{Lin2020} . In region II (blue) and for $18<\tau_{ext}<58$ nN$\cdot$\AA{},  $0<|L_1|\approx |L_2|<1$. It corresponds to a transmission of rotation between the master and the molecule-gear. It is the so-called driving regime.  Finally in region III (red)  for $\tau_{ext}>58$ nN$\cdot$\AA{}, we always found $|L_1|<|L_2|$. As the master torque is continuing to increase, $|L_2|$ is reducing further. This corresponds to an average angular velocity of the HB-BPB molecule-gear which cannot follow the master rotation. It is the so-called overdriving regime. Let us notice that as the torque magnitude continues to increase and becomes much larger than 68 nN$\cdot$\AA{}, some atoms of the \textit{tert}-butylphenyl teeth start to dissociate from the master due to the corresponding extremely high kinetic energy reached in this condition. Here, the molecule-gear slows down, the conformation of its teeth is rather deformed and the master rotation started to be erratic.

\subsection{Phononic dissipation}

\begin{figure}[t]\centering
 \includegraphics[width=0.6\textwidth]{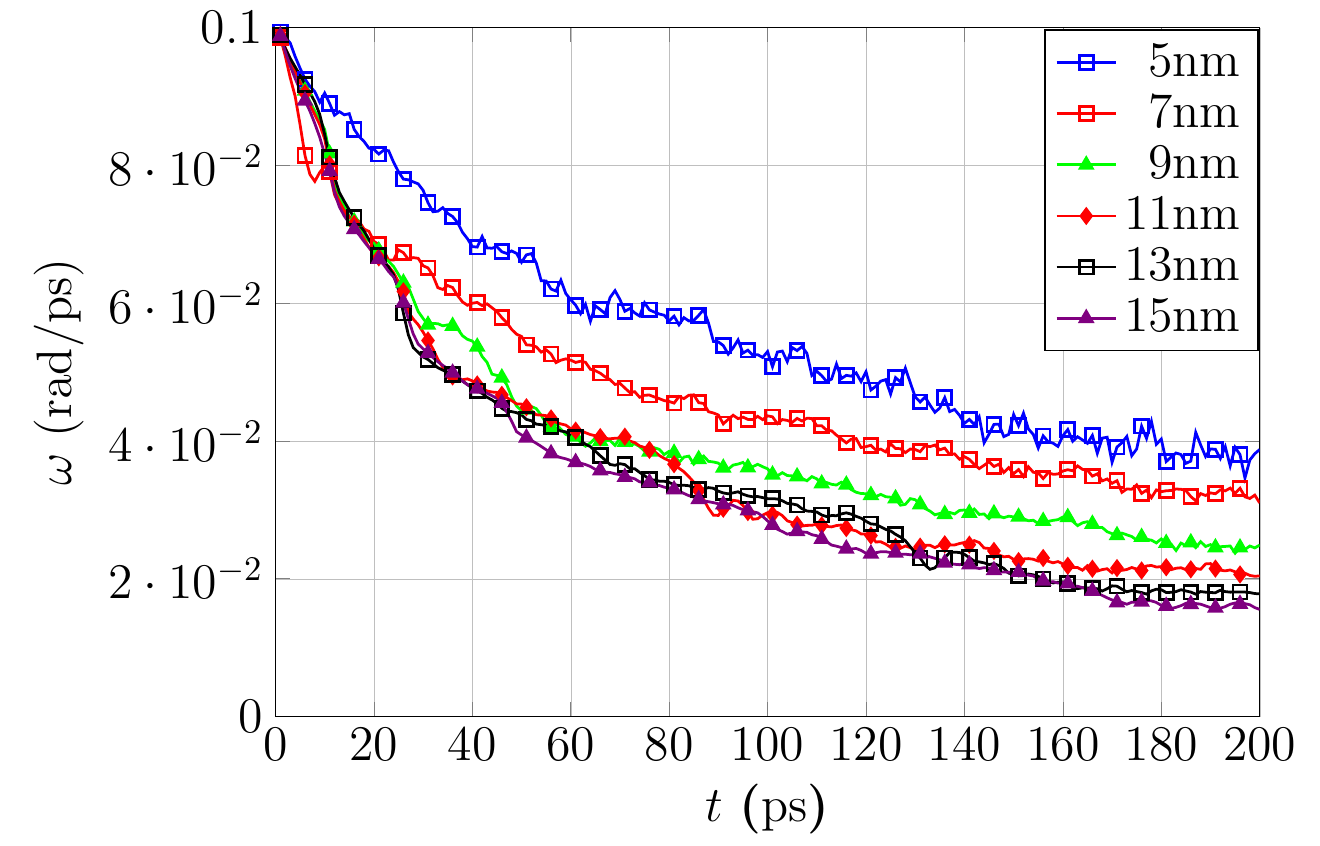}
 \caption{Viscous dissipation for graphene disks with initial angular velocity $\omega_0=0.1$ rad/ps on Cu(111) within 200 ps for different gear diameters ranging from 5 to 15 nm.}
 \label{Fig:omega_relaxation_5_15nm}
\end{figure}

In  Fig.~\ref{Fig:setup} the main friction requiring the application of a large torque is coming from the interaction between master and the Cu(111) surface (See Ref. \cite{Lin2020a}). To study the phononic friction in action, we design a numerical experiment only for the master by specifying an initial angular velocity $\omega_0=0.1$ rad/ps for this master without molecular teeth and central hole for simplicity. For different nanodisk diameters, we let this graphene nanodisk to progressively slow down and stop due to the friction. With this initial angular velocity, one can estimate that for atoms away from the rotational axle the corresponding tangential velocities can exceed 1000 cm/s, which means that they are in a high-speed friction regime\cite{Guerra2010}. In this case, they are subject to  viscous dissipation due to the collisions with the surface Cu(111) atoms which are also displaced because of the master rotation. The angular velocity decay as a function of time is presented in Fig.~\ref{Fig:omega_relaxation_5_15nm} for different nanodisk diameters. For small diameters, the decay time is longer than for larger diameters because of the large contact area with the Cu(111) surface for the latter. 

\begin{figure}[t]\centering
 \includegraphics[width=0.6\textwidth]{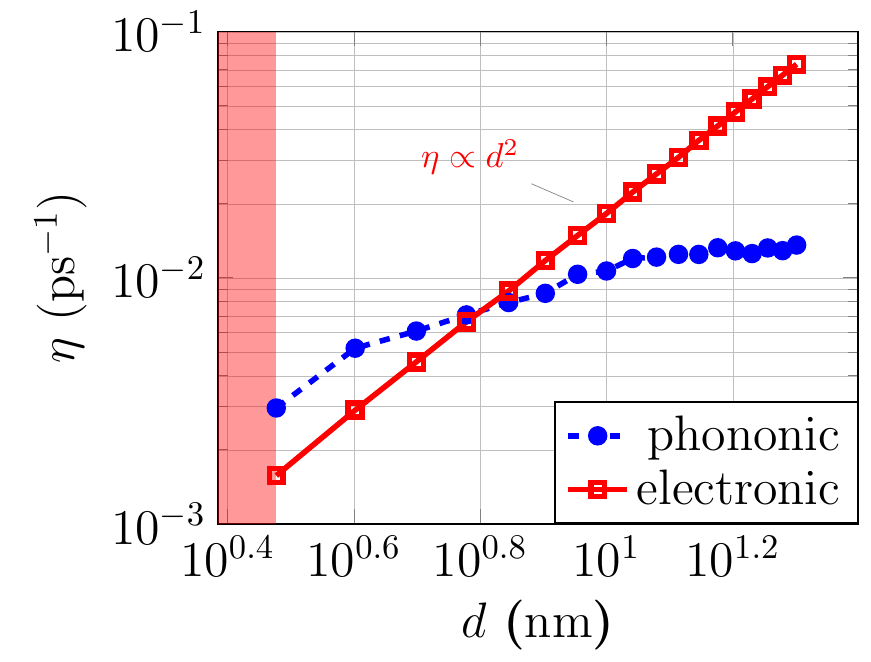}
 \caption{Size dependence of rotational friction coefficient $\eta=1/\tau$ due to the phononic contribution (blue) and the electronic contribution according to Eq.~\ref{eq:I} (red) for disk diameters $d$ from 3 to 20 nm. The red region indicates that below 3 nm the motion becomes oscillatory due to confinement by rotational barrier heights.}
 \label{Fig:relaxation_time_different_disk}
\end{figure}

\begin{figure}[t]\centering
 \includegraphics[width=0.6\textwidth]{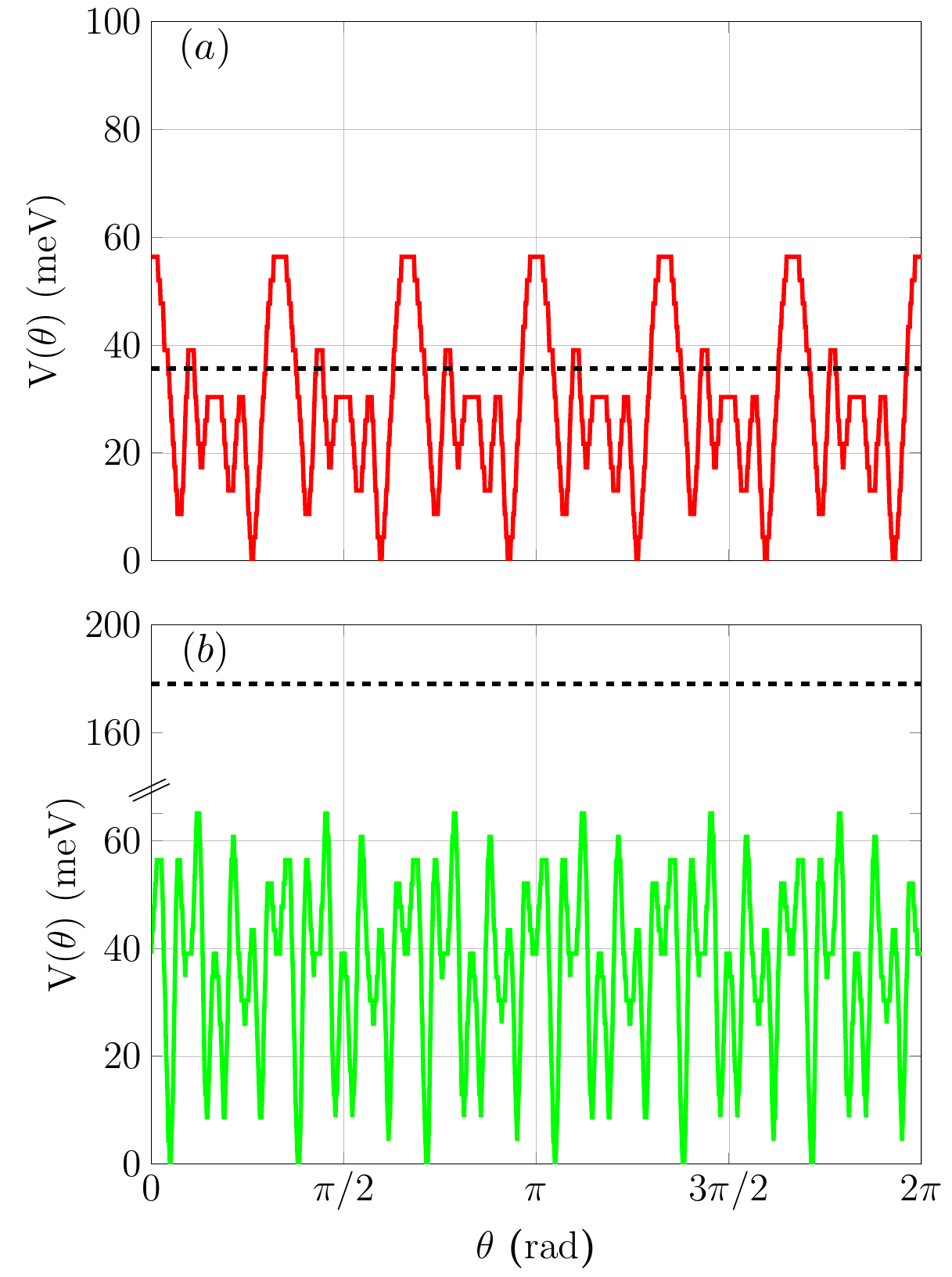}
 \caption{Profiles of potential energy surfaces $V(\theta)$ and rotational kinetic energies (black line) for (a) 2 nm and (b) 3 nm graphene disk on Cu(111) surface.}
 \label{Fig:PES_kinetic_energy}
\end{figure}

To quantify the angular velocity relaxation time, we fit its almost exponential decay with time by $\omega_0e^{-t/\tau}$. The variations of the inverse of relaxation time or friction coefficient $\eta=1/\tau$ as a function of the nanodisk diameter are presented in Fig.~\ref{Fig:relaxation_time_different_disk} (blue), for graphene nanodisk diameters between 3 nm and 20 nm. We found here with MD the friction increases as size becomes larger but it is weakly depending on size for $d>10$ nm. We notice that for small diameters ($d<5$ nm), the exponent of $d$ is 2, and this can be described by the friction theory of isolated adsorbates\cite{Persson2000,Persson1985}. The friction coefficient is given by
\begin{equation}
    \eta_{\parallel}\approx\frac{3}{8\pi}\frac{M}{\rho}\left(\frac{\omega_{\parallel}}{c_T}\right)^3\omega_{\parallel},
\end{equation} 
where $M$ is the mass of adsorbate (graphen nanodisk), $\omega_{\parallel}$ is the resonance frequency for parallel motion, $\rho$ and $c_T$ are the mass density and transverse sound velocity of the substrate. This implies $\eta\propto M(d)\propto d^2$ and is consistent with MD. However, for larger diameters the increasing of friction coefficient is suppressed since the disk is approaching an incommensurate layers with respect to substrate and creates destructive interference\cite{Persson2000}.   
On the other hand, for diameters below 3 nm (red region in Fig.~\ref{Fig:omega_relaxation_5_15nm}), the nanodisks no longer rotate but oscillate in a Brownian like motion\cite{Guerra2010}. The reason for this transition is that the potential energy barrier height for rotation on the potential energy surface (PES) describing the rotation mechanics of a graphene nanodisk on a Cu(111) surface becomes in this case smaller that the initially imposed rotational kinetic energy as presented in Fig.~\ref{Fig:PES_kinetic_energy}. For example, with a d = 2 nm nanodisk (Fig.~\ref{Fig:PES_kinetic_energy} (a)), the total rotational kinetic energy around 35 meV for an initial angular velocity $\omega_0$ = 0.1 rad/ps is below the rotation barrier height of about 50 meV. In this case, the d = 2nm graphene nanodisk oscillates randomly in its initial potential well considering also the $T$ = 10 K Cu(111) surface temperature. On the contrary, a d = 3 nm nanodisk has a corresponding kinetic energy of about 180 meV. Therefore, the rotation of the nanodisk will be weakly impacted by its mechanical rotation barrier height and its rotational motion on the Cu(111) surface will be dissipative (Fig.~\ref{Fig:PES_kinetic_energy} (b)).

%
\subsection{Electronic dissipation}
Aside from the phononic contribution, it is also important to estimate how large the electronic contribution is as compared to the phononic one active in our MD simulations. The electronic friction from the van der Waals interaction has been studied in Refs.\cite{Persson2000, Persson1995} where the metal was treated using a semi-infinite jellium model. From  second-order perturbation theory, one has the following expression for the electronic friction coefficient (in Gaussian unit):
\begin{equation}
   \eta_{el}=1/\tau_{el}=\frac{e^2}{\hbar a_0}\frac{\left[k_F^3\alpha(0)\right]^2}{\left(k_Fz_0\right)^{10}}\frac{m_e}{M}\frac{\omega_F}{\omega_p}k_Fz_0I(z_0,r_s),\label{eq:I}
\end{equation}
where $e$ is the elementary charge, $\hbar$ is the reduced Planck constant, $a_0$ is the Bohr radius, $\alpha(0)$ the static polarizability of the adsorbate, $\omega_p$ is the plasma frequency of the metallic substrate,  $k_F$ the Fermi wave vector, $\omega_F$ the Fermi frequency, $m_e$ the electron mass, $M$ the mass of the graphene nanodisk and where $I$ is the parallel frictional integral (defined in Ref. \cite{Liebsch1997}) as a function of the electron gas radius $r_s$ and of the distance $z_0$ between the graphene nanodisk and the supporting substrate (calculated from MD, this average distance is $z_0=3.1$\AA{}). For Cu(111), $\omega_p=1.33\times 10^{16}$ rad/s\cite{Zeman1987,Ordal1985}, $\omega_F=1.06\times 10^{16}$ rad/s\cite{Ashcroft1976}, $k_F=1.36\times 10^{10}$ A$^{-1}$\cite{Ashcroft1976} and $I(z_0,r_s)\approx7$\cite{Liebsch1997,Persson2000}($r_s=2.67$ for Cu). For monolayer graphene sheet, $\alpha(0)=0.949$ \AA{}$^{3}$  per  unit  cell\cite{Kumar2016}. To estimate $\alpha(0)$ for a graphene nanodisk, we assume that this disk is large enough such that $\alpha(0)\approx 0.949\times N_c/2$ (one unit cell contains two carbon atoms) where $N_c$ is the number of carbon atoms in the graphene nanodisk. Then, we can estimate the electronic contribution to the rotational friction coefficient as a function of the nanodisk diameters. The results are shown in the red curve in Fig.~\ref{Fig:relaxation_time_different_disk}.  According to Eq.~\ref{eq:I}, the electronic friction coefficient is roughly proportional to $d^2$ since the increase of polarizability is proportional to the number of carbon atoms and it grows faster than the increase of the mass of a graphene nanodisk ($\alpha(0)^2/M\propto d^2$).

In comparison with the phononic contribution to friction, the electronic contribution to friction dominates for larger disks whereas for small size disks the phononic dissipation dominates. As a consequence a large graphene nanodisk of higher polarizability could be effective in reducing electronic friction. For instance,  for other metallic substrates like Pb and Nb, the electronic friction will reduce greatly\cite{Dayo1998,Kisiel2011} when $T$ is below the critical superconducting temperature pointing out that in the  set up of  Fig.~\ref{Fig:setup} it may better to employ a superconducting surface support.

\section{\label{sec:Conclusion}Conclusion and outlook}
In conclusion, our large-scale MD simulations demonstrate that a solid state nanogear equipped with adapted molecular teeth chemical groups will be able to transfer a rotational motion to a single molecule-gear having the same molecular teeth. To fight against the surface friction of the supporting surface, a large torque has to be applied to the solid state nanogear. Experimentally and at low temperature, such a torque can be applied by one of the STM tips of a 4 STM tips instrument. But this can destabilize the chemical structure of the complete machinery in particular disengage the solid state nanogear from its atomic axle. In this prospect, experiments performed on a superconducting surface are preferable as recently performed on a molecule-gear train\cite{WeiHyo2019}. We have also assumed that the chemical structure of the solid state nanodisk is restricted to a one-layer thickness to be compatible with the effective surface height of the molecule-gear. This puts a large constraint on the solid state nanogears nanofabrication process whose thickness in our days are  in the 5 nm to 10 nm range. A 2D monolayer material like graphene or MoS$_2$ is a good starting point for nanofabricating such a solid state nanogear with a thickness in the 1 nm range which can also be modulated layer by layer. Finally, we have assumed that both the molecule-gear and the solid state nanogear are exactly rotating respecting their initial predefined rotational axles. This hypothesis requires a better exploration but depends practically on how the central axle of this solid state nanogear with be atomically nanofabricated. Low temperature STM experiments are now underway to confirm the above MD simulations.
\ack 
This work has been supported by the International Max Planck Research School (IMPRS) for ``Many-Particle Systems in Structured Environments'', the MANA-NIMS MEXT WPI program and also by the European Union Horizon 2020 FET Open project ``Mechanics with Molecules'' (MEMO, grant nr.\ 766864). We also acknowledge the Center for Information Services and High Performance Computing (ZIH) at TU Dresden for computational resources.

\bibliographystyle{iopart-num-mod} 
\bibliography{paper} 

\end{document}